# The 4-particle hydrogen-antihydrogen system revisited: twofold molecular Hamiltonian symmetry and natural atom antihydrogen

G. Van Hooydonk, Ghent University, Faculty of Sciences, Krijgslaan 281 S30, B-9000 Ghent (Belgium)

**Abstract**. The historical importance of the original quantummechanical bond theory proposed by Heitler and London in 1927 as well as its pitfalls are reviewed. Modern ab initio treatments of H-$\underline{H}$ systems are inconsistent with the logic behind algebraic Hamiltonians $H_{\pm}=H_0\pm\Delta H$ for *charge-symmetrical* and *charge-asymmetrical* 4 unit charge systems like $H_2$ and $H\underline{H}$. Their eigenvalues $E_{\pm}=E_0\pm\beta$ are exactly those of 1927 Heitler-London (HL) theory. Since these 2 Hamiltonians are *mutually exclusive*, only the attractive one can apply for stable natural molecular $H_2$. A wrong choice leads to problems with antiatom $\underline{H}$. In line with earlier results on band and line spectra, we now prove that HL chose the wrong Hamiltonian for $H_2$. Their theory *explains* the stability of attractive system $H_2$ with a *repulsive* Hamiltonian $H_0+\Delta H$ instead of with the *attractive* one $H_0-\Delta H$, representative for *charge-asymmetrical* system $H\underline{H}$. A new second order symmetry effect is detected in this attractive Hamiltonian, which leads to a 3-dimensional structure for the 4-particle system. Repulsive HL Hamiltonian $H_+$ applies at long range but at the critical distance, attractive charge-inverted Hamiltonian $H_-$ takes over and leads to bond $H_2$ but in reality, $H\underline{H}$, for which we give an analytical proof. This analysis confirms and generalizes an earlier critique of the wrong long range behavior of HL-theory by Bingel, Preuss and Schmidtke and by Herring. Another wrong asymptote choice in the past also applies for atomic antihydrogen $\underline{H}$, which has hidden the Mexican hat potential for natural hydrogen. This generic solution removes most problems, physicists and chemists experience with atomic $\underline{H}$ and molecular $H\underline{H}$, including the problem with antimatter in the Universe.
PACs: 34.10.+x, 34.90.+q, 36.10.-k

## 1. Introduction

Work on hydrogen-antihydrogen (H-$\underline{H}$) interactions remains inconclusive, despite *ab initio* techniques [1]. These theories have their roots in 1927 Heitler and London (HL) theory on HH-interactions [2a], which we review here because of its historically important connection with intra-atomic charge inversion, unknown in 1927. Using a conventional charge distribution, HL concluded that the stability of $H_2$ is due *to the anti-parallel spin-alignment of the 2 valence electrons and their exchange* [2a]. In the early 1960s, the long range behavior of spin-spin coupling, described by [2a], was criticized by Herring [2b], following earlier work by Bingel, Preuss and Schmidtke [2c] and Sugiura [2d]. Failing to solve H$\underline{H}$ is strange, since Kolos and Wolniewicz [3] proved that spectral data for $H_2$ and its PEC (potential energy curve) can be calculated accurately and since $H_2$ and H$\underline{H}$ are two 4-particle systems with similar constitution and complexity. If one system can be described exactly, the other should also be, at least if HL theory were correct and *ab initio* techniques were really reliable. After a successful description of $H_2$, Kolos and others also analyzed H$\underline{H}$ 30 years ago [4] but could not get a complete and conclusive view on the H-$\underline{H}$-interaction. Many novel attempts to further quantify system H$\underline{H}$ followed [1].

The difference between molecular systems HH and H$\underline{H}$ and between atomic systems H and $\underline{H}$, is due to a simple parity operator **P**=±1. As quantum theory easily deals with symmetry, this failure is even stranger, as it occurs for the simplest system of all, hydrogen. Failing to reach a conclusion on HH is important since recent studies on HH [1] are intensified by claims that mass-production of antihydrogen H *seems* possible [5]. The *expectation* is that the H-spectrum or its important term 1S-2S will be available soon, which is important for CPT, for the WEP, deriving from Einstein's relativity theory and for antimatter. But, like chemists, physicists also experience difficulties with H-symmetry. An exact QFT solution for H does not exist, although H is assessed with great accuracy, *theoretically and experimentally*. These uncertainties about chiral behavior of systems H, H and HH, HH are clearly exposed since experimentalists [5] must come to the rescue to settle these problems.

In the advent of Physics/Einstein Year 2005, solving antiatom H (physics) or interaction H-H (chemistry) and preferentially both is required. The real origin of the difficulties must lie in the first solutions ever for these systems, which means that Bohr H-theory (1913) and Heitler-London $H_2$-theory (1927 [2a]) must be revisited, probably along the same lines as in [2b-2d]. To reach a transparent conclusion from this historical survey, we start the analysis by focusing on Hamiltonians of 4 particle systems HH and HH, rather than on wave functions in *ab initio* methods [1-4]. The effect of symmetries *within* Hamiltonians is more important, simpler to understand and more straightforward to quantify than that of symmetries in the extremely complex wave functions of [1,4]. We show that revisiting HL-theory can lead to a simple classical but drastic solution for molecular HH, which, in turn, leads to a drastic and unconventional solution for atomic H. By extension, a problem of matter-antimatter symmetry in the Universe can be avoided.

**2. Molecular Hamiltonians and intra-atomic charge inversion**

*Wave mechanics becomes classical physics again, when the numerical value of a wave function is constant and put equal to +1*, which is useful to discuss Hamiltonian symmetries. Then, there is an inconsistency in the effect of *inter-atomic interactions* in 4-unit charge system $H_2$, when described exclusively with (a) the HL Hamiltonian *without a parity operator* and (b) the HL wave function *with a parity operator*.

With this constraint, it is evident that the discrete symmetry exhibited by natural system $H_2$, e.g. its *mutually exclusive* singlet and triplet states, must be solely attributed to wave functional symmetry: this is the classical HL-solution. But as soon as there is an overlooked Hamiltonian symmetry having the same effect, the HL-conclusion becomes doubtful, since it must be established which parity operator is at the origin of the observed splitting. *This is what happens in the case of HH and HH*, when looking back at HL-theory [2a]. The origin of 2 *mutually exclusive* states for a quantum system is a parity operator **P** but, today and with the hypothesis of charge-inversion, exactly this parity is a major problem for HL-theory [2a] as apparent with [1,4]. To prove this in detail, we start with the non-relativistic 10-term HL Hamiltonian for a 4-unit charge system (2 leptons a, b and 2 baryons A, B with lepton-baryon *charge-conjugation* for both atoms Aa and Bb) like in $H_2$

$$\mathbf{H}_+ = +\tfrac{1}{2}m_a v_a^2 + \tfrac{1}{2}m_b v_b^2 + \tfrac{1}{2}m_A v_A^2 + \tfrac{1}{2}m_B v_B^2 - e^2/r_{Aa} - e^2/r_{Bb} + (-e^2/r_{Ab} - e^2/r_{Ba} + e^2/r_{ab} + e^2/r_{AB}) \quad (1)$$

The only difference with the 1927 HL-notation is a subscript + and a collection of *inter-atomic* terms between brackets, *expected to be responsible for bond formation and stability*. HL used the *conventional* Bohr-type charge distribution, valid at the time. *The inter-atomic terms in* (1) *may seem decisive for bonding but this is not absolutely true.* The reason is that HL do not allow for asymptotic freedom for system $H_2$, since they neglected ionic structures, for which a different asymptote is required (see below). With $\mathbf{H}_+$, HL *only fixed the asymptote at the atomic dissociation limit*, since $H_2$ *normally* dissociates in 2 neutral atoms H at $r_{AB}=\infty$. But with the atomic dissociation limit shifted to the origin, it is impossible to conclude with (1) from which side this zero limit is approached by 2 interacting neutral H atoms [2b-2d]. Realizing this, it is essential to verify the character of HL Hamiltonian (1): is it generically bonding (attractive) or anti-bonding (repulsive), as in the Bingel-Preuss-Schmidtke and Herring critiques [2b,2c]? *Only if its last term* $+e^2/r_{AB}$ *for the proton-proton inyteraction were decisive in the bond formation process, the atomic asymptote is reached from the repulsive side, contradicting the essence of HL-theory on the stability of the* $H_2$ *bond. This is the main theme of our further analysis of stable molecular hydrogen, in line with* [2b,2c].

Hamiltonian (1) applies for the two *charge-symmetrical* H-H and H̲-H̲ interactions

$$\mathbf{H}_+(HH) = \mathbf{H}_+(\underline{HH}) \quad (2a)$$

an extension unthinkable of in 1927. Two charge inversions leave the sign of all *Coulomb* terms in (1) unaffected. Referenced to the asymptote, interaction $\Delta\mathbf{H}$ (usually denoted as a perturbation) is the same for HH and H̲H̲ and leads to one *covalent* bond energy $D_{cov}$ for homonuclear systems HH, H̲H̲

$$\Delta\mathbf{H} = +(-e^2/r_{Ab} - e^2/r_{Ba} + e^2/r_{ab} + e^2/r_{AB}) = -D_{cov} \quad (2b)$$

unless there would be an energy difference between atomic systems H and H̲. This question [5] is treated below but, since this difference, if it exists, is expected to be small it can be neglected in first approximation to fix the *atomic dissociation limits* for both systems H+H and H̲+H̲.

For (2b) to be bound, HL-theory implies that the following inequality holds

$$e^2/r_{Ab} + e^2/r_{Ba} > e^2/r_{ab} + e^2/r_{AB}$$

a plausible hypothesis, *but difficult if not impossible to prove or validate* [2b-2d]: in the HL-model, bonding is achieved by virtue of the extra 2 lepton-baryon attractions, created when two neutral atoms get close.

Using the wave mechanical procedure with atomic wave functions with a built-in symmetry, HL found in 1927 [2a] that the eigenvalues for natural stable molecular system $H_2$ are given by

$$E_\pm = E_0 \pm \beta \tag{2c}$$

where $E_0$ is the eigenvalue for the atomic asymptote and $\beta$ the eigenvalue of the *resonance* or *exchange* integral. *The appearance of exchange forces in HL-theory was surprising and considered as a triumph for wave mechanics, since these forces are unknown in classical physics.*

However, for *charge-asymmetrical* H-$\underline{H}$, $\underline{H}$-H systems [1,4,6], an algebraic switch (a parity operator **P**) appears. Its effect is restricted to the 4 *inter-atomic* terms in (2b) [6], since *intra-atomic* terms remain as in HL-approach (1). As only part of the terms in (1) is affected by one intra-atomic charge inversion, this gives another Hamiltonian **H** with a *different internal algebra* (symmetry) than (1)

$$\mathbf{H} = +\tfrac{1}{2}m_a v_a^2 + \tfrac{1}{2}m_b v_b^2 + \tfrac{1}{2}m_A v_A^2 + \tfrac{1}{2}m_B v_B^2 - e^2/r_{Aa} - e^2/r_{Bb} - (-e^2/r_{Ab} - e^2/r_{Ba} + e^2/r_{ab} + e^2/r_{AB}) \tag{3}$$

the starting point in all studies on H$\underline{H}$ [1,4,6]. Here, the asymptote contains neutral *atoms* H and $\underline{H}$ and if the last term in (3) were again decisive for the interaction, *the same asymptote is now reached from the attractive side, suggesting that, instead of $H_2$, only H$\underline{H}$ can be the stable molecular hydrogen system, present in nature.* This drastic solution, if valid, would solve most of the problems with $\underline{H}$ [6].

For charge-inverted Hamiltonian (3) to be bonding at the atomic asymptote, the inequality

$$e^2/r_{Ab} + e^2/r_{Ba} < e^2/r_{ab} + e^2/r_{AB}$$

must hold. This is exactly the opposite view of HL-theory *but it is equally difficult to prove or validate* and explains why 4-particle systems are insoluble. The consequences of (1) and (3) and their common asymptote $\mathbf{H}_{atom}$ will be dealt with below.

However, we see that an internal algebra for a molecular Hamiltonian is possible and that this also generates a discrete symmetry for 4 particle systems *without the use of wave functions. Obviously, this symmetry, due to charge inversion, is competitive with the symmetry in wave functions, due to lepton spin and exchange (permutation) in 1927 HL theory* [2a]. If so, we have an internal inconsistency with *2 different symmetries* applying for the same neutral 4 particle systems HH or H$\underline{H}$, since both symmetries are described with the same parity operator **P**. *This internal inconsistency is the more remarkable as the effect of lepton-spin on the total energy of a system containing the lepton is small. A charge inversion on the lepton however, changes the character of the system: it transforms from attractive to repulsive (or vice versa) with a considerable, if not dramatic effect on the energy of the total system to which the lepton belongs.* This implies immediately that, as soon as a discrete symmetry is observed for a system like molecular hydrogen, the chances to observe it experimentally will be far greater when this symmetry is due to charge-inversion, unknown in 1927, than when it is due to lepton spin-inversion. Only if these inversions were physically and/or formally degenerate, another problem emerges, with interesting prospects also (see below).

The asymptote problem referred to above relates, among others, to ionic structures. In fact, HL neglected asymptotic freedom for the 4-particle system they described: original 1927 HL-theory also disregarded *ionic structures* for molecular hydrogen. This neglect led to a long-standing rivalry between VB-(Valence Bond) and MO-(Molecular Orbital) theories. In MO-theory, ionic structures are as important as the covalent ones. In VB-theory, the ionic contribution is parameterized [6], a secondary problem we also discuss below.

**3. First order Hamiltonian symmetry for neutral 4-particle systems and HL-theory**

The sum of *intra-atomic* terms, the atomic asymptote $\mathbf{H}_{atom}$, and inter-atomic terms (2b) reduces the difference between Hamiltonians (1) and (3) in one generalized *algebraic* Hamiltonian

$$\mathbf{H}_{\pm} = \mathbf{H}_{atom} \pm \Delta \mathbf{H} \tag{4a}$$

having a built-in parity operator, due to charge-inversion only. The distinction between (1) and (3) on account of the + and – sign in (4a) is *purely conventional*, since this sign simply says how the interatomic terms in (1) and (3) have been collected. *With the above convention, it only seems the Heitler-Hamiltonian (1) is attractive, but this is not absolutely valid as we will show in detail below.*

Leaving this important problem as it is, it is immediately verified nevertheless that the eigenvalues of (4a) are

$$E_{\pm} = E_0 \pm \beta \tag{2c'}$$

using the same notation as in (2c). In essence, this means that HL had to create a rather complicated wave mechanical framework to obtain (2c), easily obtained with (4a) *without wave mechanics*.

*In fact, it is immediately verified, without calculations, that the eigenvalues (2c') of (4a) are formally the same as those of HL-theory (2c). Moreover, it is strange that this degeneracy of eigenvalues was never mentioned in [1,4] or that alternative solution (3) for molecular hydrogen was never considered. How can this ambiguity for explaining the observed splitting in natural system $H_2$ be removed? If this mathematical degeneracy of eigenvalues (2c) and those of (4a) is not resolved properly, this leads to the many fundamental problems, referred to above.*

*The simplest unscientific way to get a solution for this dilemma is to ignore it. In fact and historically, this is exactly what happened. Indeed, the conventional and now standard solution is extremely drastic, persists for decades but, in reality, cannot even be validated: it simply forbids (3) in nature and promotes HL-theory to the status of being absolutely valid for molecular hydrogen. This solution, adhered to by the physics and chemistry communities at large, generates problems with the presence of antimatter in the Universe, with atomic antihydrogen H̲ and with molecular HH̲ in particular. It is therefore necessary to verify, as soon as possible, if this conventional decision, i.e. forbidding (3) for natural $H_2$, is scientifically sound.*

At this stage, the unavoidable conclusion of discrete symmetry (4a) *due to charge inversion*, deriving from charge conjugation or particle-antiparticle symmetry **C**, is that Hamiltonians (1) and (3) are simply *mutually exclusive*. Only many years after Dirac particle-antiparticle theory, atom-antiatom studies on the basis of (3) were started [1,4]. It is easily verified that the above consequence of symmetry (4a) was overlooked, when all 4-particle systems HH ($\underline{HH}$) and H$\underline{H}$ ($\underline{H}$H) were studied, probably due to the *involuntary* neglect of (3) and (4a) in 1927 HL-theory [2a]. But the logic of (4a) secures that, if *one* Hamiltonian (1) or (3) in pair (4a) gives *attraction* for a neutral 4-unit charge system, the *other* must give *repulsion* for the very same system, since **P**=±1: Hamiltonians (1) and (3) being *mutually exclusive*, a basic property of quantum states for a system is obtained with (4a) without ever speaking about wave functions. At large $r_{AB}$, constant asymptote $\mathbf{H}_{atom}$ suffices for 2 non-interacting neutral atoms, *charge-inverted ($\underline{H}$) or not (H)*. For *the hydrogen species*, the absolute value for $\mathbf{H}_{atom}$ or eigenvalue $E_0$ is $2R_H = 2*109678,7737$ cm$^{-1}$ or 1 a.u. The well depth below the atomic dissociation limit is the bond energy $D_{cov}$, equal to 38292,8 cm$^{-1}$, exactly the term appearing in (2b), which will decide *by experiment* about the character of Hamiltonians (1) and (3).

*Strictly spoken, 4-particle systems are insoluble and approximations must be used to arrive at a conclusion for (1) and (3). Different models will be typified by different asymptote-specific interactions, inventoried below. The problem now is: do different justifiable approximations or models exist, other than HL-choice on the basis of (1)? If so, these should be studied closely to see what their physical or chemical implications are.*

**4. Different asymptotes for different particle aggregates and their interactions**

Not only by common sense but also by abundant evidence provided by molecular PECs, which are always written in terms of $E(r_{AB})$, the inter-baryon separation $r_{AB}$ is the decisive parameter to describe the stability of 4-unit charge systems (Born-Oppenheimer approximation). By this common sense criterion, HL version (1) is repulsive (anti-bonding) and only the charge-inverted Hamiltonian (3) can be attractive (bonding). To quantify this important common sense conclusion, we collect 8 terms in (1) and (3), allow for another constant asymptote $\mathbf{H}_{Coul}$ and rewrite (4a) as

$$\mathbf{H}_\pm = \mathbf{H}_{Coul} \pm (e^2/r_{ab} + e^2/r_{AB}) \tag{4b}$$

This does not alter the total energy of the system, only an asymptote shift is imposed, *a constant*, equal to $\mathbf{H}_{Coul} - \mathbf{H}_{atom}$. At the pure Coulomb asymptote in (4b), the separations between all 4-unit charges are infinite. If charges attract exclusively 2 by 2 in separate lepton-antilepton and baryon-antibaryon systems, *without lepton-baryon interactions being allowed*, the 4-particle system will lead to annihilation with Hamiltonian (3), since there is no repulsion term in the bound state of (4b), to prevent the collapse (*annihilation*) of these two 2-particle systems. With HL-choice (1), this same pure Coulomb system is always at the repulsive side of its asymptote $\mathbf{H}_{Coul}$. In addition, the asymptotes for the 2-lepton system $1/r_{ab}$ must separate from that of the 2-baryon system $1/r_{AB}$, on account of the mass difference. The energy set free by the attractive interaction (4b), due to (3), can be as large as $2m_ec^2 + 2m_pc^2$ (lepton-pair and baryon-pair annihilation). This is much larger than with chemical interactions, where annihilative interactions are obstructed by the (many) repulsive terms in (1) and (3) and where (angular) velocities are about $\alpha c$ instead of c. Despite this, model (4b) is of Dirac-type with **C**-symmetry, as it can account for the observed annihilation of particle-antiparticle pairs. In this case, this would proceed through the intermediary of annihilating *positronium* ($r_{ab}$) and *antiprotonium* ($r_{AB}$) systems, in agreement with observation. These systems derive from the charge distribution obeying (3), not from HL-option (1), obviously of repulsive type for this same system. *For annihilation to take place*, it is necessary that the asymptote is reached from the lower attractive side with $-1/r$, in line with Coulomb's law but which is impossible with (1).

Asymptotic freedom for the 4-particle is an essential element in the discussion of 4 particle systems on the basis of (1) and (3). *If (1) is repulsive and (3) attractive, it is difficult to accept that the nature of the 2 Hamiltonians will be inverted when the 4 particles are grouped in a different way, say 2 atomic systems, the HL approach or when another asymptote is used for the same system, i.e. when asymptotic freedom is allowed for.*

For instance, allowing for attractive lepton-baryon interactions leads to a lower asymptote and means that the 4 independent particles will have to be regrouped, *with at least one lepton-baryon system allowed*. Allowing 2 neutral atoms (2 lepton-baryon systems) brings in the HL atomic dissociation limit, where 4 particles are regrouped as 2 individual lepton-baryon or atomic systems.

To verify whether or not the symmetry (character) of the 2 Hamiltonians (4a) is conserved after regrouping the 4 particles and using another asymptote, we consider a third intermediate state of aggregation, with an asymptote in between $\mathbf{H}_{atom}$ in (4a) and $\mathbf{H}_{Coul}$ (4b).

Here, the 4-particle system is no longer a symmetrical pair, each with 2 particles (4=2+2), like in (4a) with two neutral atoms at or in (4b) with the 2 leptons and 2 baryons. In this intermediate state, the 4 particles are redistributed *asymmetrically* by virtue of +4=+1+3 =+3+1. In terms of particles, this means: *one composite particle, consisting of 3 sub-particles*, and *one non-composite particle (an elementary particle)* [6]. With respect to *even* systems (2+2) in (4a) and (4b), this intermediate state is *odd*. Yet, the advantage of this odd system is that common sense Coulomb law again becomes the only law needed for the interaction between *these new particles with an asymmetrical internal constitution*. By the neutrality constraint, charge conjugation **C** and the appearance of Coulomb's law, one has unit charge +1, the other -1. Odd 1+3, 3+1 combinations of 4-particle systems imply *a particle transfer* but must always obey

$$\mathbf{H}_{\pm} = \mathbf{H}_{ion} \pm e^2/r_{AB} \qquad (4c)$$

This third asymptote is perfectly allowed and leads to a different but very comprehensible type of interaction for the very same 4-unit charge system, described by HL-theory. With model (4c), it is immediately verified once again that (3) is still the bonding Hamiltonian, whereas HL-variant (1) remains as repulsive as it was at asymptote (4b): *first order Hamiltonian symmetry is conserved in an odd-even transition, the result of a particle transfer*. With the appearance of a composite particle with exactly one unit charge but containing 3 sub-particles, *one could be tempted to assign fractional charges to its 3 sub-particles*. These must be exactly +1/3 or -1/3 unit charge each, pending the total charge of the composite particle. For physicists, this would lead to the *now standard quark model* [6,7]. For chemists, this simply leads to a classical ionic model, wherein a *non-composite cation* (1 baryon) with total charge +1 interacts, through Coulomb's law, with its charge conjugated *composite anion* with total charge -1 (1 baryon and 2 leptons). In an *odd* ionic model, only Coulomb's law $-e^2/r_{AB}$, as in (4c) and extracted from (3), is needed to account for *attraction*, since $r_{AB}$ is the baryon-baryon separation. For the establishment, HL-theory is the standard theory for bond formation in which ionic structures like in (4c) for $H_2$ are neglected. Hence, ionic classical approximations to chemical bonding [6,11,15] have been neglected for many years also.

In fact, the main problem with (4c) is that a classical ionic approximation is, exactly as (4b), of *annihilative* type, since there is no repulsive term to prevent the ion-pair from collapsing (annihilating). This can be remedied in particular cases by introducing repulsion of the ion *cores* (like in Born-type potentials). These repulsive forces, related to system compressibility, vary like $1/r^n$, with n about 9 but these do not appear in the starting Hamiltonians (1) and (3). Such *core* for H is difficult to imagine but H can be compressed and expanded as well [8].

**5. Generic simplification of Hamiltonians (1) and (3) without a specific particle aggregate**

At this stage and looking at particle redistributions (4b) and (4c), HL Hamiltonian (1) is the repulsive one, whereas (3) is, *by exclusion of (1)*, the only attractive Hamiltonian available for a 4-particle system. By considering (4a-c) we run out of possibilities for regrouping the 4 particles, whereby 2 out of 3 asymptotes obey the Coulomb law but are of annihilative type. Another solution must be found, which must avoid, in a generic way, a 4-particle system from collapsing (annihilating). There is a straightforward way to do so without actually regrouping the 4 particles. An intermediate asymptote between (4a) and (4b) is also generated by regrouping *terms* in Hamiltonians (1) and (3) according to their individual *character*. This more abstract method is generic in that it disposes of the need to rearrange particles. One would simply collect all Coulomb terms in a sum equal to $\pm ae^2/r_{AB}$ *around the equilibrium distance* (to accommodate for intra-atomic terms in $r_{Aa}$ and $r_{Bb}$ too) and all kinetic energy terms in another sum, always of *repulsive* character $+be^2/r^2_{AB}$. This repulsive character of kinetic energy terms is apparent since, *around the equilibrium distance*, v will have to vary with $1/r_{AB}$ as expected from Bohr's H-theory.

In this fourth model, the bond energy of the stable system must be included in the energy gap to be covered by the attraction. This energy gap extends from a yet unknown asymptote $\mathbf{H}_K$, can include the atomic dissociation limit $\mathbf{H}_{atom}$ and must finish at the system's ground state energy, at least if the 4-particle system is bound. Therefore, this intermediate asymptote for a model, without a specified aggregation of particles can only be valid when $r_{AB}$ is close to the equilibrium separation ($r_{AB} << \infty$). This fourth solution for both (1) and (3) has asymptote $\mathbf{H}_K = \mathbf{H}_{ion} + be^2/r_{AB}$, which is a *pseudo-asymptote* as it cannot be constant. To remove this inconsistency, the repulsive term is added to the attractive term to create a new potential, consistent with constraint that asymptote $\mathbf{H}_K$ must be constant. Allowing for asymptotic freedom, this fourth generic and new variant for (1)-(3) becomes

$$\mathbf{H}_\pm = \mathbf{H}_K + (be^2/r^2_{AB} \pm e^2/r_{AB}) \tag{4d}$$

and must directly refer to the ground state of 4 particle systems, described by (1) or (3). Here, a stable structure gets a minimum automatically. *Since this is bound, this must obey (3) instead of (1)*. Even in this rather abstract method, *the first order symmetry of the molecular Hamiltonians is left unaffected: HL-version (1) remains as repulsive as before and can never reach the status of attractive.*

*The first advantage of this approximation (4d) by collecting the terms by character over approximations (4a-c) is that it is, analytically and theoretically, capable of producing directly a minimum for only one of the two systems HH or H*H̲ *it describes, since the two states are mutually exclusive.* The 4 different models (4a-d), with different asymptotes and interactions, illustrate how far one can go to find a reasonable solution for insoluble 4-unit charge systems. In first approximation and without wave functions, they allow to reduce semi-empirically (or in a phenomenological way by relying on the Bohr H-model for angular velocities) but significantly, charge conjugated Hamiltonians (1) and (3), at the basis of the HH, HH̲ dilemma. In 3 out of 4 cases, i.e. (4b-d), only charge-inverted Hamiltonian (3) invariantly leads to attraction, whereas HL-choice (1) always gives a repulsive anti-bonding system, as remarked earlier [2b-2d]. An extrapolation of this evidence to HL-asymptote (4a) suggests that stable system $H_2$ (HH in HL theory), must obey charge-inverted Hamiltonian (3) rather than HL Hamiltonian (1). This leads to

the internal inconsistency of HL-theory and of all studies [1,4], referred to above and to the problem with the degeneracy of end-solution (2c) in HL-theory and in its alternative (2c'), based upon (4a). *If first order Hamiltonian symmetry is conserved under asymptote shifts, the conclusion must be that $H_2$=H<u>H</u> instead of HH as in HL-theory. If so, one must accept HL-theory may well be wrong but the immediate compensation for accepting this bold conclusion would be that the H<u>H</u> and <u>H</u> puzzles would be solved as argued around (4a)* [6]. An extra argument in favor of (4d) is that, when looking at molecular spectra, (4d) is *the algebraic generalization* [9a] *of the original Kratzer-potential* [9b]. The fame of Kratzer's teacher, Sommerfeld, secures that HL must have known about (4d), since it was published in 1920 in the same journal as [2a]. This potential figures in the long list of potentials [6,9a,9c,10,11] proposed to account for many phenomena, including molecular band spectra, but which all take part in the search for the UEOS, the universal equation of state [8]. Kratzer's potential (4d) is better than Morse's, when it comes to rationalize the spectroscopic constants of more than 300 diatomic bonds, including prototype $H_2$ [11]. In this broader context, focusing on Hamiltonians like (4d) rather than on wave functions for the HH-H<u>H</u> dilemma is justified.

*The second advantage of generalized Kratzer potential (4d) in this analytical form is that it imposes that the observed PEC for the 4 particle system HH or H<u>H</u> will have to obey an equation of the second degree in variable $1/r_{AB}$, an amazingly simple result, easily verified with the band spectrum of the system, to which this Hamiltonian refers*, as we will show below. These bold predictions on the PEC for 4-particle system $H_2$ are simply impossible with HL-theory [2a]. Fitting the $H_2$ band spectrum in this way will, if applicable, reveal the unknown asymptote $\mathbf{H}_K$, invisible, neglected and overlooked in HL-theory. Only experiment will tell how reasonable result (4d) is for the spectrum of molecular hydrogen (see below). The outcome of this confrontation will also tell us how to interpret (2c) or (4a): either by a charge-inversion, forbidden in nature by convention, or by lepton spin-inversion as prescribed by standard HL-theory.

## 6. Antihydrogen atom <u>H</u>

For physicists, the main attention goes directly to unit <u>H</u>, rather than to the chemistry of H<u>H</u>. Chemical variants (4) reduce to 2 parity-related atoms H and <u>H</u>, *enantiomers*, generated by intra-atomic charge inversion, a consequence of **C**-symmetry. For atoms H or <u>H</u> occupying the dissociation limit $\mathbf{H}_{atom}$ in (4a), atomic level energies *are expected to be invariant to this internal intra-atomic charge-inversion*. With principal quantum number n, Bohr theory implies that, with reasonable accuracy (order 0,01 cm$^{-1}$ or 10$^{-6}$ eV), the identity

E(n)(H) = E(n)(<u>H</u>)                                                                                                   (5)

holds, which fixes $\mathbf{H}_{atom}$ in (4a) for all possible H+H, H+H̲, H̲+H̲ and H+H̲ interactions. With reasonably accurate (5), one cannot distinguish between these asymptotes, due to the uncertainty about the energy of H and H̲. However, *identity (5) is disproved as soon as a left-right energy difference between H and H̲ exists but it is expected that, when it exists, it will be small, without significant repercussions for all chemical asymptotes* like $\mathbf{H}_{atom}$ in (4a). Finding out how large this parity violating energy difference is, is the ultimate goal of [5], as it is important for CPT and for Einstein's WEP (see Introduction).

A chemical solution for H̲ is nevertheless obtained as soon as it can be proved that (3) is the *exclusive* Hamiltonian for natural stable system, molecular hydrogen. Unlike HL-theory, this system should be regarded as a HH̲ system, meaning that charge-inversion is allowed in nature, contradicting conventional belief. Consequently, if species HH̲ really exists in nature, the 2 *mutually exclusive* forms (enantiomers H and H̲) of species hydrogen are tolerated in nature too.

*And if molecular band spectra provide evidence for the existence of HH̲ in nature, atomic line spectra will have to provide signatures for the reality of the two states H and H̲ in natural species hydrogen. These two types of signatures must be found, since they are complementary. The detection of one type of signature, say of molecular type, does not even make sense without the detection of a signature of atomic type (and vice versa),* as remarked previously [12].

What does this mean for physics (H and H̲) and for chemistry (HH and HH̲)?

For physics, *expectation* (5) is difficult to prove, since, in first order, the Hamiltonians for H and H̲ are invariant to an *intra-atomic charge inversion*. In reasonably accurate Bohr theory (with an atomic wave function equal to +1) and *leaving out recoil*, classical atomic Hamiltonian

$$\mathbf{H} = \tfrac{1}{2}mv^2 - e^2/r \qquad (6)$$

(without wave functions) applies indeed to both H and H̲ and is at the basis of (5). With this accuracy (errors of 0.01 cm$^{-1}$), *Bohr theory cannot generate with (6) eigenvalues for left- and right-handed states of hydrogen* H and H̲. Despite its relative accuracy, Bohr's formula $-R_H/n^2$ is *achiral*: it cannot account for *chiral behavior* in atomic hydrogen, if it existed [8]. Today, spectral accuracy is not a problem. The real problem is mass-producing H̲ [5] to prove or disprove (5), since physicists *are convinced* that H̲, just like all antimatter, is forbidden in the natural world. But if observed molecular spectra are compatible with HH̲, observed atomic spectra will have to be compatible with H̲. For atomic physics, line spectra can reveal exactly those *chiral symmetry breaking terms* missing in the original *achiral* Bohr Hamiltonian (6). If these can be retraced, these results run ahead of the ongoing antihydrogen experiments [5], as argued in [12].

For chemistry, the parity operator in all models (4) proves that it must be relatively easy to detect the difference between *charge-symmetrical* system HH and *charge-asymmetrical* HH̲, *if it existed*. If it proves difficult to study unit H̲ on account of (5) and as proved *de facto* by [5], it may be easier to verify how H will interact, using the approximate schemes (4) above. The question is: is a composite system HH or HH̲ either *attractive* (bound, stable) or *repulsive* (anti-bonding, unstable)? HL-theory *seems* conclusive but it is not, due to its inconsistency with all other and valid models (4b-d), as argued above. All studies [1,4] adhere to the HL-model. To arrive at a conclusion, *only the first order symmetry (character) of Hamiltonians (1) and (3) can reveal the symmetry of the systems they stand for*. Knowing that *2 mutually exclusive Hamiltonians* (1) and (3) exist by virtue of charge inversion, we must first of all be absolutely certain about their *character: only one can be attractive, the other must be repulsive*. This fundamental problem is left out in [1,4] since *only the attractive Hamliltonian can apply to natural and stable system, conventionally called the hydrogen molecule* [6].

## 7. Hamilton character (or symmetry), the direct consequence of charge distributions

After 1930 Dirac-theory, the first question for chemists should have been if HL-choice (1) was really the best and the only one possible for the 4-unit charge system with 2 neutral atoms hydrogen. Why was variant (3), with its opposite character (symmetry) inspired by Dirac-theory, not considered immediately? For reasons difficult to understand in an historical perspective, the HL-choice (1) is still accepted today, almost like an international standard, for neutral 4 particle systems and remained so for over 75 years, despite the possibility that charge-inversion in Hamiltonians is, at least, *a theoretical alternative for wave function based symmetries*, see (2c) and (4a).

In fact, with generally accepted quantum HL-theory for 4 particle systems, **H₊** applies for *charge-symmetrical* H-H or H̲-H̲ interactions (parallel dipole alignments ↑↑ and ↓↓ in terms of *charges*), which automatically leaves **H₋** for *charge asymmetrical* H-H̲ or H̲-H interactions (anti-parallel dipole alignments↑↓ and↓↑) as it should [1,4,6]. With (4a), *charge-symmetrical interactions* HH and H̲H̲ must always separate from *charge-asymmetrical interactions* HH̲ and H̲H *in a discrete way* for any value of $r_{AB}$, just like the (lepton spin-based) singlet-triplet splitting in molecules [6], according to HL-theory [2a] and as criticized in [2b,2c]. But HL had to use wave functions with built-in symmetries exactly, as argued above, *to remedy for the repulsive character of Hamiltonian (1) they used to describe molecular hydrogen*. This led to an anti-parallel spin alignment for valence electrons (leptons) and to composite wave functions to account using *repulsive* (1) for *attractive* and *stable* system $H_2$. To the best of my knowledge, HL-theory was never seen in this important historical context.

Most physicists as well as chemists seem to be unaware of the generic implications of charge inversion leading to (4), since (3) is still forbidden for natural systems. One could have done easily the exercise like we suggested in 1985 [13], *purely out of theoretical interest*, with Hamiltonian (3) for system $H_2$, with as unique benchmark, Kratzer potential (4d), available since 1920. If this exercise had been done properly and in due time, *it may well have had a considerable effect on the further developments in theoretical chemistry as well as in atomic physics, where it led to bound state QED*. But since this exercise was not done at the time it was most needed, the original 1927 HL-approach [2a] is still considered today as a masterpiece of quantum mechanics and highly praised as such in many textbooks. It remains *the most important contribution to the theory of the chemical bond in the 20$^{th}$ century* [14]. The new insight it provided for *covalent bonding* in system $H_2$ by means of *exchange forces*, not having an equivalent in classical physics, is still the main argument in its favor. Unfortunately [6], it was also the final blow for classic ionic bonding proposed in the early 19$^{th}$ century, despite the common sense of the fundamental nature of the Coulomb law, at its basis [6,15]. *In reality, this is a fatal and historically important misjudgment, as is easily proven mathematically.*

First, let us use an earlier argument. Although ionic model (4c) was immediately abandoned after HL-theory, it is known to perform well for many stable 4-particle systems, i.e. *non-covalent*, polar or heteronuclear molecules, whereby attraction $-e^2/r_{AB}$ is prominent [6,11]. Due to (4a), *HL (1) can never be mathematically* inverted from its generic repulsive character $+e^2/r_{AB}$ to an attractive character $-e^2/r_{AB}$. At the very best, *a pseudo-ionic HL approach is possible using the escape method by regrouping particles in the odd ionic way*, discussed above. But *in a pseudo-ionic HL-model, ions can never attract directly by* $-e^2/r_{AB}$ but only *indirectly* by virtue of the 2 *lepton-baryon interactions* $-e^2/r_{Ab} - e^2/r_{Ba}$, exactly as argued already in 1985 [13].

Second, one should have been suspicious on the real message behind the mathematical equivalence of spin- and charge-operators: apart from scale factor 2, they are identical. Is this accidental or has something been overlooked? Is the degeneracy reported above of (2c) and the eigenvalues of (4a) at the origin of these 2 mathematically equivalent yet physically different operators? *Why can or should two different approaches exist simultaneously to explain the stability of the same natural system molecular hydrogen, although one predicts HH, the other H<u>H</u>?* Spin- and charge-effects upon *the total energy of a system* being completely different, *drastic effects in (bonding) energy like in (4a) are therefore always in favor of (forbidden) charge-inversion rather than (conventional) spin-inversion.*

Third, even common sense tells that HL-model (1) is *not the attractive Hamiltonian*. To imagine how atomic dipoles interact, one can rely on a classical well-known example: the interaction of 2 permanent linear magnets. A parallel configuration (↑↑ or ↓↓) is always repulsive and unstable. It can only be transformed in a stable system by a *permutation* of one of the two magnets (↑↓ and ↓↑) [6]. This simple verifiable experiment would lead to the generic conclusion that *charge-asymmetrical*

Hamiltonian (3) is *bonding*, which is immediately contradicted by HL-theory, where *charge-symmetrical* (1) must be bonding. Yet, the HL- *proof* is much more complicated than the simple verifiable experiment with 2 permanent magnets (dipoles).

Fourth, *a permutation of charges is a charge-inversion* or *a charge-exchange*, to use the HL-terminology. The HL-permutation proceeds through the wave function. With a single atomic product wave function

$1s_A(a)1s_B(b)$

no stable system can be obtained in HL-theory [2a], see also [2b,2c]. Only allowing a permutation (an exchange) as well as a symmetry with atomic *composite* wave function

$1s_A(a)1s_B(b) \pm 1s_A(b)1s_B(a)$

leads to a stable bonding (singlet) and an unstable anti-bonding (triplet) state with Hamiltonian (1), itself devoid of the corresponding algebraic switch. With (3), the permutation with respect to (1) proceeds directly in Hamiltonian, not in the wave function, which explains the degeneracy above.

Fifth, a different argument to flaw the HL-approach is provided by the prospect of annihilation, as argued for (4b-c). Due to Dirac particle-antiparticle theory and the detection of annihilating particle pairs, one invariantly *expects* that H-$\underline{H}$ interactions (3) lead to *annihilation* [1,4,5]. Is this *expectation* met or not? Here, common sense learns that, for annihilation to be *possible at close range* (small $r_{AB}$), the energy $E(r_{AB})$ of *charge-asymmetrical* system H-$\underline{H}$, must first go the more stable *attractive* side of its asymptote **H**$_{atom}$ in (4a), much like a *singlet* state of a bond in HL-theory. And when (3) is *exclusively* connected with *an attractive* (singlet state) system *on account of (4b-c) and the prospect of annihilation*, HL Hamiltonian (1) can only apply for a *repulsive* system (triplet), as argued in [2b,2c]. This elementary consequence of annihilation is also in contradiction with HL-theory, stating that only *charge-symmetrical* systems H-H ($\underline{H}$-$\underline{H}$) with (1) lead to *bonding* (attraction) as in H$_2$.

Since HL theory is the basis for all *ab initio* studies on H$\underline{H}$ [1,4], the same basic inconsistency applies for all these approaches too. With HL-theory, **H** (3) *cannot be bonding (attractive), although with the prospect or expectation of annihilation, with the experiment with the 2 magnets, approaches (4b-d)..., it should.*

Analytically, this is easily confirmed when looking at the more important terms in (1) and (3). If $r_{AB}$, the inter-baryon separation of the 4 particle system is really decisive for the stability of a structure containing these two building blocks, it is evident that $+e^2/r_{AB}$ in (1) leads to repulsion and forbids annihilation [2b,2c]. In 1962, Herring wrote explicitly: '*Closer inspection of the Heitler-London calculation for the hydrogen molecule shows that at very large interatomic separations, it becomes physically unreasonable and its predictions impossible* [2c]. *For at large $r_{AB}$, the positive $1/r_{AB}$ term in (1) becomes dominant* [2d], *being larger than* *the other terms…'* This dominance of the resulsive term $1/r_{AB}$ in (1) was exposed already in 1927 also by Sugiura in the same journal [2d].

With these views, (3) containing $-e^2/r_{AB}$ becomes attractive by definition, but the *conditio sine qua non* for this to be true is that this term is indeed the most important and decisive one for the formation of a stable 4 particle system. Reminding the mutually exclusive character of the 2 Hamiltonians in (4a), it suffices to prove analytically the absolute character of only one Hamiltonian of the two. With the analytical results of [2b-2d], (1) is proved repulsive at long range, which makes (3) attractive by definition. *Although this is also simple common sense*, it must be verified more thoroughly than in [2b-2d], especially at shorter range. Here, the problem with the exclusive attractive character of (3) must be solved in an irrefutable way. Despite the vain attempts in [1], problems are solved immediately if a stable bonding intermediate, non-annihilating H$\underline{H}$-complex existed and could be observed by spectroscopic methods (see below).

**8. The Universal Equation of State (UEOS), Hamiltonian character and attraction by -1/r**

Discussing a system solely with its Hamiltonian is in itself ambiguous. Using a wave function exactly equal to a constant +1 in wave mechanics means that classical physics is reproduced, the viewpoint of the classical 19$^{th}$ century physicist. A physicist educated with wave mechanics would think that, if this method works, some *universal kind of wave mechanics* exists, where *the numerical value of any wave function of wave mechanics* can be replaced by +1. *This means that, in essence, the wave function is of secondary importance*. This bold conclusion is confirmed by illustrious examples, of which the most obvious is, again, simple system H. Its energy levels obtained with Hamiltonian (6) are identical, whether one uses Bohr physics *without wave functions* or Schrödinger physics *with wave functions*. For H, the analytical properties of the wave function are such as to not contradict the Bohr energy result. In addition, their analytical properties are also such as not to contradict term splitting, not accounted for by Bohr theory, but by Sommerfeld's additional or secondary quantum number $\ell = n-1$, originally arrived at *without wave functions* too. Up to this level of accuracy for the theory, *wave mechanics* as such has no added value for H. A wave function *allows for corrections needed to accommodate fort shortcomings in a Hamiltonian, thought to be sufficiently accurate to describe a system* (a good example is bound state QED). Claims that wave mechanics has *ab initio* status like in [1] is an overestimate, as proved by Kolos' work on both systems H$_2$ [3] and H$\underline{H}$ [4]. Only for H$_2$, *spectral data are available*, which led to very good results by *selecting exactly these wave functions, capable of reproducing the spectral data with (1)*. But when *these same good wave functions* are applied to (3) and H$\underline{H}$, *for which no spectral data exist*, uncertainty prevails. In some of the works on H$\underline{H}$ [1], *rather subjective criteria are used for selecting so-called well-performing or well-behaving wave functions.*

*If quantum mechanics were really reliable and absolutely valid –as argued in Dirac's time -, the antihydrogen problem should not even exist, given the simplicity of its structure and of the parity operator involved.* Dirac once wrote that "…the underlying physical laws necessary for the mathematical theory of a large part of physics and *the whole of chemistry* are thus completely known…[but that the]…equations [are] much too complicated to be soluble" [16]. For chemistry, Dirac clearly referred to HL-theory [2a]. Looking at (2c) and (4a), his conclusion on *chemistry* could well be wrong, pending the tests with (4d) below. For physics, Dirac-based bound state theory for atom H was flawed by the Lamb and Retherford experiments. Surprisingly or not, the Lamb shift is directly linked with a signature needed to prove the presence of $\underline{H}$ in nature, as stated in 2002 [12] (see further below).

*Despite these flaws, theoretical and computational chemists are still convinced that Dirac was right and still believe that the only correct solution for atom-atom interactions is provided with HL-theory and its many modern variants* [1,4]. This shows why an objective analysis of (4a), and the degeneracy of its results with (2c), is required as soon as possible. And to do so, the importance of the attractive term $-1/r_{AB}$ for the stable 4-particle structures must be proved beyond doubt.

One of the main characteristics of almost all attempts to find the UEOS is that attraction by $-1/r$ is their most important common element: *only this is exclusively connected with (3) and mathematically impossible with (1)*. As in Coulomb's law, attraction by $-1/r$ is a common sense approach, an idea that goes back to Newton's times. For a variety of phenomena in many body systems, ranging from macroscopic changes in the state of aggregation to microscopic BEC transitions…*attraction by $-1/r$ is the rule, never the exception* [8]. A classical Hamiltonian approach with a leading attractive term in $-1/r$ is the only one that fits in the long list of attempts to find the *universal equation of state* (UEOS) or, in the present terminology, the *universal Hamiltonian*. This equation, if it existed, should account *analytically* for any phase transition for any system on whatever scale (micro or macro) [8]. Many scientists try to find out, most of the time *empirically*, how this intriguing UEOS should look like analytically [6,10,13,17]. For the universal chemical bond or 4-particle system under discussion here, the UEOS is a *universal potential*, which directly gives a universal numerical PEC for any molecule and for any type of bond. Elsewhere [13], we showed that the Kratzer function (4d) is a serious candidate to be of universal type and it may even open the way to find this intriguing UEOS.

With this broader context of the UEOS and the fundamental character of Coulomb's law, all evidence points towards the decisive role played by $-1/r_{AB}$ for attraction and, hence, for stable systems. If so, the attention must focus on Hamiltonians rather than on wave functions. And if it turns out that wave functions are needed with a discrete symmetry, this is simply a message *that a discrete symmetry (a parity operator) is missing in the Hamiltonian used for the system. Using the UEOS and Coulomb's law as measures, Hamiltonian $H_-$ (3) is bonding by definition, which makes HL-option $H_+$ (1) repulsive also by definition. This definitely settles this question about the symmetry (or character) of mutually exclusive Hamiltonians for 4-particle system (1) and (3) and the stable 4-particle system they stand for: by virtue of (4a) and (2c), molecular hydrogen must be denoted as charge-asymmetrical H$\underline{H}$ not as charge-symmetrical HH* [6].

### 9. Second order Hamiltonian symmetry. Analytical proof for $H_2 = H\underline{H}$

Before analyzing experimental data, we already presented enough evidence to conclude that the generic character of charge-inverted Hamiltonian (3) is attractive, given its first order symmetry implications. In addition, a secondary symmetry effect due to charge inversion exists, which is in favor of Hamiltonian (3) as this effect is excluded for *HL-option* (1). Additional secondary symmetry, implicit in *charge-asymmetrical* scheme (3), is only valid *for lepton-baryon terms* in (3) but not in (1). *Again, without using wave functions of any kind or combinations thereof*, terms in (3) can also be reordered as

$$H_- = +\tfrac{1}{2}m_a v_a^2 + \tfrac{1}{2}m_b v_b^2 + \tfrac{1}{2}m_A v_A^2 + \tfrac{1}{2}m_B v_B^2 + (-e^2/r_{Aa} - e^2/r_{Bb} + e^2/r_{Ab} + e^2/r_{Ba}) - e^2/r_{ab} - e^2/r_{AB} \qquad (7)$$

wherein *lepton-baryon Coulomb interactions* are collected between brackets. With (7), the effect of this second order symmetry for lepton-baryon interactions in 4-particle *structure* H$\underline{H}$ becomes evident. In fact, this opens the door for *only one critical geometrical arrangement of the 4-particle H$\underline{H}$ complex, for which all 4 lepton-baryon terms vanish exactly 2 by 2 -- a theoretical possibility excluded if HL Hamiltonian (1) is used*. With this secondary symmetry element, bound state $H_-$ (7) is also simplified considerably (from 10 to 6 terms) but in a different way [6]. Although a minimum in (4d) is easily obtained by putting its first derivative ($d/dr_{AB}$) equal to 0 (see below), (7) throws another light upon the collection of terms by character and on the spatial or geometrical particle arrangement in this approximation. For this critical spatial configuration, (7) reduces to

$$H_- = (+\tfrac{1}{2}m_a v_a^2 + \tfrac{1}{2}m_b v_b^2 - e^2/r_{ab}) + (+\tfrac{1}{2}m_A v_A^2 + \tfrac{1}{2}m_B v_B^2 - e^2/r_{AB}) \qquad (8a)$$

$$[= \quad \text{positronium} \quad + \quad \text{antiprotonium}]$$

by virtue of the very stringent secondary symmetry requirement for (7)

$$(-e^2/r_{Aa} - e^2/r_{Bb} + e^2/r_{Ab} + e^2/r_{Ba}) = 0 \qquad (8b)$$

Here, (8b) is an extra stability or symmetry constraint for 4-particle systems, not yet discussed above and exclusively connected with charge-inverted Hamiltonian (3). By this extra condition, a stable neutral 4 particle system is generated, consisting of 2 neutral subsystems *positronium* and *antiprotonium*, whereby the position of the positronium system is either symmetrical (*covalent bonding*) with respect to the two charge conjugated baryons in antiprotonium or asymmetrical (*ionic bonding*). *In wave mechanics, wave functions are measures for spatial system configurations (representations). But, once again, this is not the case with generic equation (8), which imposes a stringent analytical condition for the configuration of a 4-particle system directly in the Hamiltonian, a condition impossible to reach with continuously varying wave functions.*

Moreover, critical symmetry (a configuration) (8b) for complex HH̲ is automatically and always obtained with a *charge-inverted ionic system* (ion anti-ion pair), only *possible with a particle transfer* (odd system +4=+1+3=+3+1). *In this case all 4 lepton-baryon terms in (7) always cancel exactly and can be disregarded in the attractive Hamiltonian (3)* [6].

Despite its simplicity, (8b) is a rather drastic criterion indeed, *since it disposes of the effect of the 4 lepton-baryon interactions in the bound 4-particle system along Coulomb field axis $r_{AB}$* but these lepton-baryon interactions are exactly the terms needed in HL-theory to explain covalent bond formation and stability. *Without attractive lepton-baryon terms, HL-theory could not even survive*, as shown above. The possibility that lepton-baryon interactions could disappear from the scene by virtue of simple geometric symmetry effect (8b) can be interpreted *as if leptons and baryons, known to interact strongly at long range, suddenly and at short range did no longer interact and were completely free to move within the 4-unit charge system.* This interpretation of secondary symmetry (8b), together with the asymptote shifts and the idea of fractional charges +1/3 and –1/3 discussed above, brings in the essentials of the quark model for particle physics. In this way, molecular hydrogen can be interpreted indeed as a 4 elementary particle system, but only on account of (8b), which means *that molecular hydrogen be denoted by HH̲*.

==========================================================
INSERT FIG. 1 and FIG. 2 AROUND HERE
==========================================================

Critical geometries (structures) are shown in Fig. 1 and 2 [6]. These are also needed to explain the rather subtle difference between classical ions and charge-inverted ions. Fig. 1 gives *pseudo-ionic* models in HL-model (1), whereas Fig. 2 gives the corresponding really ionic structures with charge inverted Hamiltonian (3). Completely or 100% ionic structures with shape ├ or $A^-B^+$ and ┤ or $A^+B^-$ are not shown, since the underlying interaction details cannot be displayed properly.

Fig. 1 shows clearly why bonding *lepton-baryon attractions* are vital for HL-theory. The classical *covalent* HL-model, Fig. 1b, in between two *pseudo-ionic structures* 1a and 1c, has become an attractive positronium antiprotonium model in Fig. 2b. This can be seen as a 19[th] century mechanical Watt regulator or a gyroscope. The static baryon system in antiprotonium is under control of a *rotating* positronium system (in a plane perpendicular to $r_{AB}$) and situated exactly in the middle of $r_{AB}$ in the case of a homonuclear bond like $H_2=H\underline{H}$. It is easily verified by graphical inspection of Fig. 2 that, whatever the position of the positronium system with respect to the baryons, the projection of the lepton-baryon interaction on the inter-particle field axis ($r_{AB}$) is always zero, as required by symmetry (8b). For heteronuclear ionic as well as for anti-ionic systems (*with a permanent dipole moment*), the 2 leptons will be displaced towards the more *electronegative* atom.

These are the concrete 3-dimensional structures generated for the bound 4-particle system, just on the basis of *the twofold Hamiltonian symmetry* discussed above, *without using any wave function*.

It is evident from (8a) that, after collecting and, eventually disregarding lepton-baryon terms by the second symmetry constraint (8b), the result of (8a) after collecting terms of the same character is again a Kratzer-potential like in (4d).

However, there is a significant difference: although a difference may occur for the value of coefficients a and b as well as for the value of the Kratzer asymptote $\mathbf{H}_K$ in (4d), the Kratzer variant generated with (8a) *can only be bonding, as repulsive Coulomb terms are excluded by* (8b). With constraint (8b), (4d) can only transform in the original *attractive* Kratzer-potential of 1920

$$\mathbf{H}_{-} = \mathbf{H}_K + (be^2/r^2_{AB} - e^2/r_{AB}) \qquad (8c)$$

*which is the only attractive solution left for a 4 particle system obeying (3). This equation finally proves analytically that Hamiltonian (3) is the only one possible to lead to a bond for molecular hydrogen, which cannot be denoted but by* $H\underline{H}$. Hence, focusing on Hamiltonian symmetries proves to be quite productive *historically, conceptually and even analytically*.

Although for (4d), it was difficult to say something about critical particle configurations, this uncertainty is now removed, as illustrated in Fig. 2, due to the secondary symmetry (8b), applicable for (3) but not for (1). Exactly as for (4c), *ionic charge-asymmetrical* $H^+\underline{H}^-$ or $H^-\underline{H}^+$ configurations reduce the 6 terms in (8) to the 2, already given in (4c) [6].

Schematically, covalent molecule $H_2$ can be considered as referring to *ionic* structures [11,15]

$$H_2 = H\underline{H} = \tfrac{1}{2}\,[(H^+\underline{H}^- + H^-\underline{H}^+] \qquad (8d)$$

If wave functions are necessary to describe covalent bonding at close range, ionic rather than atomic wave functions will have to be used for Hamiltonian (3). This is completely different than the procedure applied in [1,4] and sheds a new light on the difference between VB- and MO-theories [6]. *On the basis of this twofold symmetry of the molecular Hamiltonian (4a) and (8b), completely absent in original HL-theory, there is only one solution possible for a stable 4-unit charge system: charge-inverted Hamiltonian (3) is exclusively connected with bound stable systems, whereas HL Hamiltonian is doomed to be repulsive. If so, neutral composite antiparticles must be allowed in nature, and by extension, also antimatter. This contradicts the conventional* *solution for (4a), by excluding charge-inversion in natural systems, as argued above. The original Kratzer potential (8c) is and remains the generic bonding solution for (3) and, as such, must apply to a bound neutral 4-particle system. Solution (8c) must be confronted with experiment for molecular hydrogen, the only stable and natural 4 elementary particle structure yet assessable by spectroscopic means.*

## 10. Identifying the stable hydrogen-antihydrogen complex: the natural $H_2$ molecule

To verify the attractive or repulsive character of Hamiltonians (1) and (3), the only reliable and objective source of information is the spectrum of a 4-particle system, a chemical bond. This will reveal the shape of its PEC, its well depth and the position of its minimum. Only the PEC of a 4-particle system can disclose the exact path followed by its sub-systems when these interact and form or do not form a stable system. Spectroscopy is a very efficient, if not the only available, tool to decide about the applicability of our final result (8c) as well as of potentials (4a-d) and to remove the dilemma on the interpretation of eigenvalues (2c) and those of (4a). Before going into the important details, we give a global view on the four approximations (4a-d), which includes potential (8c).

*10.1 PECs for all systems (4a-d) and (8c)*

Generic results (4a-d), based upon an asymptote shifting procedure, are presented graphically in Fig. 3a and 3b. Instead of $r_{AB}$, number $n=r_{AB}/r_0$ is used as a numerical variable, the *reduced* inter-baryon separation. For pure Coulomb systems, we apply $r_0=1$ Å, but for $H_2$-related PECs, $r_0=0,74144$ Å. Linear variable n is used for all PECs in Fig. 3a, whereas for Fig. 3b inverse $1/n$ applies. Table 1 gives all quantitative data for models (4a-d), used for the construction of Fig. 3.

Table 1 Potentials and asymptotes used for all PECs in Fig. 3a and 3b

| # | Equation | $r_0$(Å) | Asymptote (au) | Potential | Remarks |
|---|---|---|---|---|---|
| 1 | (4b) | 1 | 1 | $(1\pm 2/n)$ | 2 Coulomb systems (+ branch not shown) |
| 2 | idem | 1 | 1 | $(1\pm 1/n)$ | 1 Coulomb system (+ branch not shown) |
| 3 | (4c) | 1 | ½ | $½(1\pm 1/n)$ | ionic system (+ branch not shown) |
| 4 | (4d),(8c) | 0,74144 | ½ | $½(1\pm 1/n)^2-0,18$ | 0,18 au $H_2$ bond energy (part of + branch) (8c) is bound state of (4d) |
| 5 | (4a) | 0,74144 | 0 | $+0,025/n$ | HL, guess for repulsive (2b), partly shown |
| 6 | | 0,74144 | 0 | ? | RKR PEC for $H_2$ |

==========================================================
INSERT FIG. 3a AND FIG. 3b AROUND HERE
==========================================================

As a reference point, the PEC for $H_2$ is also given with a minimum at 0,74144 Å and well depth of –0,17447 a.u. (see further below). It is situated at the attractive side of the repositioned atomic dissociation asymptote 0 in (4a). This global view on all approximations for Hamiltonians (1) and (3) is essential to distinguish clearly between them with as sole and decisive reference: the observed band spectrum of natural molecular hydrogen [6].

Let us start with the seemingly worst approximations (4b) with two variants (cases 1 and 2 in Table 1) and (4c) in both Fig. 3a and 3b. Their repulsive branches, due to HL Hamiltonian (1), are not shown in order not to lose the details of their attractive branches in the bonding region, i.e. around asymptote 0 and further below. Despite the simplifications to arrive at simple Coulomb potentials (4b-c) for which $r_0$=1 Å is used (see Table 1), they nevertheless all end in the critical bonding region of natural system $H_2$, either directly at its minimum: (4b), version 1-2/n or at the intersection with asymptote 0: (4b), version 1-1/n *and* ionic model (4c). In this respect, also the ionic potential (4c) behaves properly in the critical region, despite its total neglect in HL-theory. The apparent convergence of so-called bad or naïve Coulomb potentials (see Table 1) is surprising, especially since they all use Hamiltonian (3) and are impossible with HL Hamiltonian (1). The picture is simpler in version 1/n in Fig. 3b. Here, linearity allows extrapolation of the Coulomb inverse power law to two different worlds (+ and -), a combination mathematically forbidden for inverse power laws [6]. This seemingly bad behavior of Coulomb approximations for system $H_2$ is easily removed, since the real universal properties of the Coulomb law for 4-unit charge systems like $H_2$ are easily exposed [6]. For the assessment of HL-theory, the results are amazing when looking at the analytical behavior of simple *ab initio* Kratzer potential (8c), i.e. $\frac{1}{2}(1-1/n)^2$, when compared with the observed PEC, see Fig. 3a. With its limitation to the bonding region, it is only natural to see that it diverges from the PEC at the extremes (see Fig. 3b). The approximation used for HL-result (4a) in Table 1 may seem suspicious but, to the best of my knowledge, a better approximation for (2b) *without the use of wave functions* is not available (at least, none was retrieved in the literature).

Both Fig. 3a and 3b show that in the critical region at long range, *a multitude of attractive potentials, all deriving from (3)*, must cross the single repulsive one, generated by HL Hamiltonian $H_+$. Exactly here, interference of long-range forces (dipole-dipole, Van der waals interactions... [8]), cannot be excluded. But this must not alter our conclusions based upon (1) and (3), wherein interactions of type $1/r^n$ with n>2, are absent. The presence of a repulsive part in the PEC at long range (before the critical distance is reached), as shown in Fig. 3a and b, would lead to small maximum (an instability region) for the neutral 4-particle system. If a (small) the maximum is detected, this long-range part of PEC, can, in our approach, only occur due to the repulsive HL Hamiltonian $H_+$ (1) [2b,2c]. This maximum disappears when the HL PEC would cross any of the PECs, generated by bonding Hamiltonian $H_-$. Long-range maxima in PECs are known for long (see Fig. 3 in Varshni's review [10]). Of course, up-down behavior of molecular PECs (i.e. up at large separations, down at small range) is a signature for a phase transition. If so, the $\mathbf{H}_+,\mathbf{H}_-$ distinction (4a) may be important for other phenomena too, and its use must not be restricted to chemical bonds. Ultimately, this brings in a discussion about the UEOS and phase transitions, an important issue as argued above but not further discussed here [8].

Instead, given the good behavior of the Kratzer potential in Fig 3, we now confront in detail the observed $H_2$ spectrum with a second degree fit, imposed by the Kratzer potential (8c).

*10. 2 PEC for natural stable system: the hydrogen molecule, confronted with Kratzer potential* (8c)

The detailed $H_2$ PEC $E(r_{AB})$ is shown in Fig. 4a. To illustrate the amazing power of the generic Kratzer potential (see Fig. 3), we remind that the PEC generated by the original HL-method [2a] had a similar shape as that in Fig. 4a but, in terms of accuracy for well depth and position of the minimum, it was rather bad. It took James and Coolidge 6 more years to calculate a better one [18] and even 30 more years were needed for really good results [3]. But using the same technique for H<u>H</u>, these same authors were uncertain about the PEC for system H<u>H</u> of interest here [4]. The RKR PEC for $H_2$ used in this work for Fig. 3 and 4 is taken from [19].

================================================================
INSERT FIG. 4a AND FIG. 4b AROUND HERE
================================================================

A difficulty with the E, $r_{AB}$ presentation in Fig. 4a is that, despite its smooth form, it is difficult to fit, suggesting that $r_{AB}$ is not the best variable. In fact, it is contrary to expectation for Hamiltonians (4b-d), all suggesting, like the UEOS, a more natural inverse $1/r_{AB}$ dependence, even for repulsive states. Fitting the curve in Fig. 4a with a $4^{th}$ or $6^{th}$ order polynomial leads to bad results.

Therefore, the simple $2^{nd}$ order fit, imposed by (8c) for the bound state, is applied to the more natural E, $1/r_{AB}$ presentation (Fig. 4b). This must lead, given the fundamental nature of (8c), to a value of Kratzer asymptote $\mathbf{H}_K$. The generic value chosen in Table 1 corresponds with $1R_H$, 0,5 au or the ionization potential of atom H, $IP_H$. Despite the simplicity of (8c) compared with HL-theory, it gives an acceptable picture for the observed PEC (see Fig. 3a and 3b).

For the fit, the highest levels as well as zero and first level are disregarded, as indicated in Fig. 4b. There are various reasons to legitimate this procedure. First, the asymptote for (8c) is not HL-asymptote (4a) and since (8c) like (4d) is confined to the minimum, the highest levels can be omitted. Next, there is an uncertainty about the inversion procedures for constructing PECs. Different competing methods (Rydberg-Klein-Rees or RKR method, Inverted Perturbation Aproach or IPA,…) exist, the details of which are not discussed here [6,13,19]. This is why the first level is skipped also, since inversion discrepancies can be large in the neighborhood of the minimum. Using $r_{AB}$-values in Å for the turning points in Fig. 4a and 4b allows a direct quantification of coefficients a and b with a 2$^{nd}$ order fit. One of these is Kratzer asymptote $\mathbf{H}_K$, obtained at $1/r_{AB}=0$ or $r_{AB}=\infty$ (see Fig. 4b), where it intersects the axis. For the remaining 9 levels (18 data points), the goodness of a simple 2$^{nd}$ order fit is acceptable ($R^2 = 0,99$). The result is

$$E(1/r_{AB}) = 38595,4/r^2_{AB} - 109147,7/r_{AB} + 78394,9 \quad cm^{-1} \tag{9}$$

which, to make sense, must be interpreted in terms of *available atomic or molecular constants* of H and (8c). With the H-electron affinity $EA_H$ equal to 6081,4 cm$^{-1}$ and $IP_H$ equal to the Rydberg, the absolute position of the ionic asymptote $IP_H+EA_H$ is at +115760 cm$^{-1}$. If this gap is shifted towards the minimum of the well depth $-D_{cov} = -38283$ cm$^{-1}$, its new position is at +77477 cm$^{-1}$. With fit (9), the asymptote for Kratzer potential (8c) is at +78395 cm$^{-1}$, a difference of only 1.2 % with the value derived with the constants.

Using the information on the constants, the coefficients in (9) for system H$\underline{H}$ can all be identified as

$$E(1/r_{AB}) = D_{cov}/r^2_{AB} - IP_H/r_{AB} + (IP_H+EA_H-D_{cov}) \quad cm^{-1} \tag{10a}$$

*This is an amazing and unexpected result for theoretically predicted (4d)* and impossible with the repulsive Hamiltonian $\mathbf{H}_+(1)$ of HL-theory. The Kratzer asymptote $\mathbf{H}_K$ in (8c) is positioned at

$$\mathbf{H}_K = (IP_H + EA_H) - D_{cov} = \mathbf{H}_{ion} - D_{cov} \tag{10b}$$

This means that the gap, covered by Kratzer potential (8c), is exactly the same as the gap between the ionic asymptote $\mathbf{H}_{ion}$ in (4c) and the atomic one $\mathbf{H}_{atom}$, appearing in HL-theory (4a) [6]. In fact, in the classical ionic approximation (4c) for a hydrogenic system

$$\mathbf{H}_{ion} = IP_H + EA_H \tag{10c}$$

is simply the energy of anion H$^-$ but also of charge-inverted anti-anion $\underline{H}^-$. With charge-inversion, composite cations with total unit charge +1 cannot be excluded.

The equilibrium constraint by taking the first derivative of (8c) gives an equilibrium separation for the baryons, equal to $r_e = 2*38595,4/109147,7 = 0,71$ Å, close to the observed value (0,74144 Å). As remarked before [6], generic result (8c) may help to improve current inversion techniques.

The generic approach on the basis of twofold Hamiltonian symmetry not only solves the H$\underline{H}$ problem. It immediately leads to unprecedented results regarding the stability of the 4-particle system, not even imaginable in the context of HL-theory. *Asymptotic freedom for chemical systems* implicates that the HL-asymptote is nothing else than a trompe-l'oeil (*an optical illusion*) [11]. In other words, the stability of molecular hydrogen must be explained with the charge-inverted ionic asymptote, for which (3) applies, instead of with the atomic asymptote of HL-theory. Exactly HL-theory is at the basis of the rejection of ionic bonding, suggested in the early 19$^{th}$ century by people like Berzelius [6]. This historical mistake must be corrected by 2005, as exactly this rejection led to unnecessary problems with antihydrogen.

*The most important result, however, is that the dilemma about the interpretation of eigenvalues (2c), the result of standard HL-theory with wave functions but also the generic result of the charge-inverted Hamiltonian without wave functions, is solved definitely in favor of charge-inversion symmetries, conventionally forbidden in nature.*

Given the importance of this conclusion, one should verify if results (10) in favor of (3) are not accidental and exclusively applicable for simple system HH. This confirmation can be achieved by studying other 4-particle systems (chemical bonds).

*10.3 Confirmation from 37 other bonds based the lower order molecular spectroscopic constants*

A consistency check using PECs for many other bonds is elaborate [6]. To avoid the inversion procedure for PECs, working directly with observed molecular constants is possible by using a very elegant method proposed already 50 years ago by Varshni [10]. In this method, observed first order molecular spectroscopic constant like $\alpha_e$ (a first order rotational constant) and $\omega_e x_e$ (a first order vibrational constant) suffice to compare directly the spectroscopic characteristics of bonds (4-particle systems), including their PECs and their properties around the equilibrium distance (the well depth). The Varhsni method [10] was used in previous work [11]. Varshni drew the attention to the Sutherland parameter $\Delta$ [20], called so by him in honor of Sutherland. For 4-particle systems (bonds), $\Delta$ combines 3 major equilibrium properties

$$\Delta_{cov} = \tfrac{1}{2} k_e r_e^2 / D_{cov} \qquad (11)$$

since $k_e$ is the force constant, $r_e$ the equilibrium distance (inter-baryon separation) and the asymptote, in this case $D_{cov}$, the covalent bond energy, all important parameters for a band spectrum. The obvious connection with HL-theory (1) is the use of atomic asymptote $D_{cov}$. With Kratzer-type results (8c) and (9)-(10) deriving from (3), the better asymptote would be $\mathbf{H}_{ion}=D_{ion}$, the ionic bond energy $D_{ion}$ instead of the covalent bond energy $D_{cov}$.

=============================================================I

INSERT FIG. 5a AND FIG. 5b AROUND HERE

=============================================================

When Varshni function F, analytically related to rotational constant $\alpha_e$ [10] is plotted versus $\Delta_{cov}$ for 39 diatomic bonds or 4-particle systems, including H$_2$, a single straight line is expected theoretically [10,11,21]. Fig. 5a shows the actual result (this figure is reproduced from our 1982 paper [21]). The curves shown in Fig. 5a are: curve *ion* derives from a simple ionic Born-Landé potential, curve *Morse* is the prediction of the Morse potential and curve *cov* refers to an empirical equation due to Varshni [10]. To arrive at a Sutherland parameter, it must be realized that $k_e$ as well as $r_e$ are determined by experiment and *that only the choice for the asymptote is free*. Fig. 5a clearly reveals that a choice for HL-asymptote $D_{cov}$ in (11) leads to *a large spectroscopic gap between ionic and covalent 4-particle systems*, difficult to understand if HL-theory was universally valid, i.e. for any type of 4-particle system or bond. HL-related choice (11) generates a large and fundamental difference between *covalent and ionic bonding*, invisible in either Hamiltonian (1) or (3). The conclusion from Fig. 5a is that the use of $D_{cov}$ advocated by HL-theory, *is not really of universal character, since it does not apply to all types of bonds* [11,21]. *Hence, HL-theory cannot be the universal theory either, despite convention and despite its general acceptance by the establishment.*

As shown in this work by (10c), deriving from (3), an ionic more universal Sutherland parameter

$$\Delta_{ion} = \tfrac{1}{2} k_e r_e^2 / D_{ion} \qquad (12)$$

should provide a better solution [21]. Plotting F versus $\Delta_{ion}$ gives Fig. 5b (also reproduced from [21]). The gap of Fig. 5a, due to (11), has simply disappeared, even for the homonuclear or covalent molecules including $H_2$, analyzed in detail above [21] (the only *aberration* is $F_2$).

A completely similar situation is found for Varshni function G, related to vibrational constant $\omega_e x_e$ for the same 39 bonds (see [21] for the details).

The confirmation needed for the result of the foregoing paragraph, stating that stable molecular hydrogen must be denoted as H<u>H</u>, is provided by 38 other bonds or 4-particle systems, all more complex than simple prototype $H_2$. This validates our interpretation of the eigenvalues (2c) as being due to charge-inversion as well as the reliability of Kratzer potential (8c).

Looking at these large scale results (39 bonds), it goes without saying that HL-theory based (11), fails exactly for those bonds it was meant to describe so accurately: *covalent* 4-particle systems like $H_2$, $Li_2$, $Na_2$, $K_2$… As argued above, this failure of HL-theory is due to the neglect of asymptotic freedom for 4-particle systems and of internal Hamiltonian symmetries (4a). *The rejection of ionic bonding by the chemistry establishment was indeed a fatal, if not an historical misjudgment.*

Moreover, we tested ionic asymptotes and the Kratzer potential with the spectroscopic constants for more than 300 chemical bonds or 4-particle systems [11]. This elaborate study confirms the usefulness of (12), and therefore the universal properties of the Kratzer model (8c) and results (10). With some modification, the method was astonishingly accurate and led to a variety of conclusions, we cannot all repeat here. For instance, we argued, with Varhsni [10], that *the famous Dunham expansion* [22], considered as a standard for molecular spectroscopy, must be rejected also as it uses the wrong variable for the inter-baryon separation $r_{AB}/r_0$, whereas, as shown above, the inverse Kratzer variable $r_0/r_{AB}$ is the better choice [11]. The argument here is again common sense. Kratzer potential (8c), even when expanded, will always converge, as its character is attractive, like that of (3) from which it derives. Due to the analytical form of its variable, the Dunham expansion can never converge but is nevertheless used persistently for bound systems for many a decade. The analogy with wave mechanics is evident: to arrive at a reasonable convergence with the non-converging Dunham potential, the number of Dunham-coefficients needed is almost infinite. Even then, only moderate convergence can be achieved. Similarly, using a repulsive Hamiltonian (1) to describe a bound system, like Heitler and London did, the number of wave functions needed to achieve (only moderate) convergence will have to be almost infinite too. In theoretical or computational chemistry, just like in [1], extremely complicated wave functions, even with more than 100 terms, are the rule rather than the exception.

Again, all this evidence suggests that, to explain bonding at short range even for molecular hydrogen, one should, if any are needed, use ionic wave functions of type *multiplicative* type $1s_A(a)1s_A(b)$ and/or $1s_B(a)1s_B(b)$. Additive ionic functions like $1s_A(a)1s_A(b)\pm 1s_B(a)1s_B(b)$ will secure the total system does not exhibit a dipole moment, see (8d). We also verify that the already proven repulsive behavior of the HL-procedure at long range [2b-2d] remains valid even at shorter range (see below).

*10. 4 Additional evidence from atomic and molecular constants*

Additional consistency checks of alternative bonding scheme (3) for 4-particle systems are easily made and go beyond the simplest systems HH and H<u>H</u>. A *chemical* check is obtained by equating the 2 descriptions of the total well depth for any covalent or homonuclear 2-atom system $X_2$ [11]. The first is the sum of *ionization potentials* $IP_X$ and *covalent bond strength* $D_{cov}(X_2)$, which refers to the HL atomic dissociation limit. With an ionic dissociation limit, *electron affinity* $EA_X$ and *ion separation* $r_{XX}$ are also needed. Since the two methods describe the same asymptote difference (the total well depth), the identity $2IP_H+D(H_2)=IP_H+EA_H+e^2/r_{AB}$ leads, amongst others, to results of type

$$e^2/r_{AB} + EA_H = IP_H + D(H_2) \qquad (13)$$

which can easily be tested with experimental data for atoms other than H. This simple result refers in its own way to the degeneracy of eigenvalues (2c) and those of (4a).

==============================================================
INSERT FIG. 6 AROUND HERE
==============================================================

With this degeneracy, it appears that even for so-called insoluble 4 particle systems, molecular and atomic data are very simply correlated as in (13). Fig. 6 illustrates the validity of (13) for 12 for *homonuclear* bonds $X_2$ with X=H, Li, K, Na, Rb, Cs (series □) and Ag, Au, F, Cl, Br, I (series ○). Despite the simplicity of argument (13), it is obeyed very neatly, although the data separate in two sets, referring to the position of elements in the Columns of the Table [11].

*10. 5 Critical distance*

Using a similar argument, it is also straightforward to estimate the critical distance where the transition from Hamiltonian (1) to (3) may occur. This is easily visible with Fig. 3a and 3b where crossing of various curves at long-range is illustrated. This critical distance for two atoms H obeys

$$e^2/r_{crit} = IE_H - EA_H \qquad (14)$$

The critical region is of interest to compare VB- and MO-theories but also to test HL-theory. Particle transfers at the critical distance are relatively easily to observe with femtochemistry [23]. As before [6], we again associate this critical distance with the transition from an atom-atom HL system (1) to a charge-inverted atom-antiatom system, obeying (3), where a Kratzer or Coulomb potential takes over (see Fig. 3a and 3b), if crossing is avoided [6,11].

*10. 6 Consequence from molecular evidence on natural chemical system HH for atom system hydrogen*

It is remarkable that the very same 19$^{th}$ century common sense classical ionic Coulomb interactions (4c), abandoned in favor of *exchange forces* soon after HL-theory was published in 1927 [2a], must now come to the rescue to solve the HH-problem. In fact, the interference of ionic interactions, starting at an asymptote between 0 and $2*IP_H$, explain why the apparatus of wave mechanics is so complex. It was once laughingly said that *quantum chemistry cannot but weigh a captain of a ship by weighing the ship when is and is not on board* [24a]. With this metaphor, ionic interactions find their origin within the ship as its asymptote is well above the atomic one, i.e. within the interior of the ship.

More generally, computational problems with modern ab initio quantumchemistry, using HL Hamiltonian (1), MO with full CI, are established beyond doubt and proven *de facto*. Some of these can be avoided with a less stringent DFT but the main cause for these computational difficulties is commonly known as the *Coulomb-problem*, e.g. the repulsive leptonic term $+e^2/r_{ab}$ in Hamiltonian (1). Modern codes rely on the possibility to cut-off (part of) this annoying repulsive term, see for instance [24b]. With charge-inverted (3), lepton repulsion becomes attractive, which seems to offer an alternative to deal with this long standing Coulomb-problem (electron correlation), implicit with (1). But it must be evident by now that the natural stable HH bond has been identified as the well-known hydrogen molecule, usually denoted by $H_2$. As a condition sine qua non, this immediately implies that equally simple spectroscopic signatures should exist to prove the presence of charge-inverted H, i.e. antihydrogen H̲, in nature too. This brings in atomic physics and the spectroscopic identification of natural atomic species H̲ (see Introduction).

**11. Confirmation by signatures for natural H̲ in the available H-spectrum**

Exactly as with the abundant molecular spectroscopic evidence above for the identification of HH̲ as simple natural $H_2$, but persistently overlooked for about 75 years, a similar almost identical situation applies for the identification of natural atom H̲. A first misjudgment was made about the reliability of ionic bonding and natural charge inversion in the case of chemical bonding but a second similar misjudgment was made in atomic physics many decades ago about the reliability of the Bohr Hamiltonian (6), producing reasonable results for atom H, without using wave functions.

As we stated before [12], *solving chemical problem HH̲ only makes sense, if the presence of H̲ can also be proved in nature.* The strange thing is that, just like we had to use 19$^{th}$ century chemistry to solve the problem with HH̲ in nature, we are again forced to use 19$^{th}$ century evidence to prove the reality of H̲ in nature too. This evidence relates to *the discovery of chiral behavior* (mirrored structures, enantiomers, optical activity…) and its importance in nature by pioneers like Pasteur, Van't Hoff, Lebel…

Hund set out the constraints for chiral behavior [25], in the same year and in the same journal as HL-theory [2]. Hund found that the PEC for a system, able to manifest itself either as a left- or a right-handed structure, cannot consist of a single well but must have two wells. One well is for the left-handed, the other for the right-handed structure, with a maximum in between the 2 wells. These PECs are of Mexican hat-type. Given this evidence, it suffices to find a Mexican hat type PEC within the observed spectrum of natural system H. If this can be detected, the existence of natural enantiomers H and H̲ is proved.

The generic approach to left-right transitions is provided with a 3-dimensional Cartesian reference frame. The mathematics is simple if the symmetry is not violated. The possibility that this symmetry is violated, leads to the difficulties in physics and chemistry. Nevertheless, the mathematical model leads to some basic generic characteristics or signatures for a transition from a left- to a right-handed structure (reference frame). It is indeed possible to quantify the most important signatures a priori or in a generic way since these signatures are system independent. In fact, 3 generic signatures are available to detect the presence of H̲ with spectral data. These are rather elemental but all are subject to only one constraint

1. if H and H̲ are really mirror images, the mirror plane must be situated at 90° or ½π radians
2. if hydrogen exhibits chiral behavior, a double well or Mexican hat PEC must exist

3. if a permutation (inversion) is at work, the end result is a rotation by 180° or π radians

4. common constraint: *these 3 signatures are not covered by simple Bohr atom theory, due to (5).*

Condition 4 is proved since Bohr H-Hamiltonian (6) does not contain any term to represent chiral behavior or an internal parity operator related with such behavior. In fact, a charge-inversion or a permutation of charges (signature 3) leaves the Bohr Hamiltonian invariant. Therefore, if signatures 1-3 exist, they can only be retraced *in the deviations of Bohr theory from experiment*, due to constraint 4. *Then, if hydrogen shows chiral behavior not covered by Bohr theory, the errors of Bohr theory cannot be at random but must show a definite very specific pattern*, which may reveal the 3 signatures even *quantitatively*, *pending the accuracy of the spectral data.*

Exactly because of its simplicity, Bohr theory can be tested in a number of different equally simple ways. One test is to verify if the hydrogen Rydberg $R_H$ is really constant, as claimed by Bohr with his famous formula for H-level energies

$$E_{nH} = -R_H/n^2 \tag{15}$$

where n is principal quantum number, giving, *without wave functions*, the eigenvalues of the atomic H Hamiltonian (6). With (15), the values of $-E_{nH} \cdot n^2$ provide with the real Rydberg or the $R_H(n)$ value needed to reproduce the exact energy for each level n, if Bohr's version of (15) were not correct. This simple analysis was actually done a few years ago for the H Lyman series [12]. $R_H$ as conceived by Bohr for (15) is not a constant at all. Its variations (the errors of Bohr theory) follow a parabolic law in function of $1/n$, with a maximum Rydberg-value of 109679.3522 cm$^{-1}$ at principal quantum number n=½π, exactly the value expected for chirality signature 1 [12]. This result was left unnoticed since the time of Bohr (1913). Signature 1 for the presence of <u>H</u> in nature is thus confirmed.

In turn this harmonic Rydberg [12] is a perfect input-value to recalculate level energies with Bohr formula (15) and to subtract these results from the observed ones, giving differences $\Delta E_{nH}$. Also this analysis was done in 2002 [26] but published in 2004 [27]. A plot of differences $\Delta E_{nH}$ versus $1/n$ produces a perfect Mexican hat or double well potential for natural system H. This result confirms Hund-based chirality signature 2 to prove the existence of enantiomers H and <u>H</u> in nature. The maximum between the 2 wells of the Mexican hat PEC is situated exactly at n= π [26,27], which confirms chirality signature 3 and proves that a permutation of charges (a charge-inversion) occurs when *natural* left-handed H goes over in *natural* right-handed <u>H</u>. Reminding the transition from molecular Hamiltonian (1) to (3), *this is an intra-atomic charge inversion*, indicating that, if left-handed H has charge distribution (+;-), right-handed antihydrogen <u>H</u> must have inverted charge distribution (-;+).

Since these 3 signatures are not covered with Bohr theory, this confirms condition 4 and fits in (5). Having found these 3 essential generic signatures for the presence of atomic H in nature proves the simple logic and the common sense of our approach towards antiatom H. In turn, these 3 signatures confirm the complete analysis above on the chemistry of HH-interactions and the validity of molecular Hamiltonian (3), despite its rejection by the establishment. All this proves the usefulness of our starting hypothesis: to focus on Hamiltonians rather than on wave functions and our restrictions about the relative *predictive* power and/or reliability of wave mechanics in general.

In addition, since the H-line spectrum shows a transition governed by a parity operator, not contained in the Bohr Hamiltonian (6), the terms responsible for the chiral behavior must be identified and incorporated in a symmetry adapted atomic Hamiltonian.

The phenomenological, semi-empirical analysis in [12,26,27] shows that the eigenvalues (15) must be adapted as

$$-E_{nH} = (R^*_H/n^2)[1 - A(1 - \tfrac{1}{2}\pi/n)^2] \qquad \text{cm}^{-1} \qquad (16)$$

with $A \approx \sqrt{\pi}$ [12,27,28] and where $R^*_H$ is the harmonic Rydberg at $n = \tfrac{1}{2}\pi$ [12].

The analyticity of chiral eigenvalues for system hydrogen (16) is quite particular in that, exactly as in the case of the chemical bond HH and HH studied above, a wrong asymptote choice can also lead to a wrong interpretation of the line spectrum. For instance, using the classical Bohr asymptote of 109678,77 cm$^{-1}$ in (16) instead of the harmonic Rydberg $R^*_H$ of 109679,35 cm$^{-1}$ [12,26,27] will distort the double well PEC in [26,27] to leave only one well, as illustrated in detail elsewhere [8]. Using a wrong asymptote for H, would suggest that H is a one well system like in Bohr-theory and illustrated with (5), whereas, in reality, it is a double well system, which explains its chiral behavior.

To order $1/n^4$, result (16) is very similar to standard bound state QED for H, based on the Dirac-Sommerfeld equations [28]. However, the critical n-value in QED for the Lyman-series is $n=3/2$, close to but different from $n=\tfrac{1}{2}\pi$ [12]. Dirac bound state theory predicted that the H np-series had to be degenerate with the ns-series, as both series were subject to critical $n=3/2$. As remarked above, this prediction by Dirac-based bound state theories, considered as absolutely valid at the time (before 1947), was flawed with the discovery of the now famous Lamb shift [29]. The standard Lamb shift can therefore be connected with the difference between $3/2$ and $\tfrac{1}{2}\pi$ [12] and, by virtue of signature 1, with chiral behavior in natural system H. It appears that, historically, a unique occasion to prove the reality of H was wasted more than 50 years ago when the Lamb shift was discovered. Rather than focusing on the possibility that chiral behavior could be involved, this discovery led to the increasing complexity of bound state QED, as we know it today [28] but which, because of our results, cannot yet be validated [12]. It is remarkable that an *unexpected* byproduct of the present analysis is a striking similarity between this *classical model* for 4 elementary particle systems (bonds) and the *quark model* for (many) elementary particle system in general.

## 12. Conclusion

When looked at in an historical perspective, the main problem with H̲ is that there should not be a problem at all. It is time to end the speculations, dreams and phantasies on H̲ by physicists: patents on H̲-production and storage [30] as well as the prospect of using H̲-driven vehicles for deep space travelling, even appearing in refereed physics journals [31].

It appears that the reliability, robustness and simplicity of 19$^{th}$ century Hamiltonians and their *mutually exclusive* character (bonding or anti-bonding, attractive or repulsive) for neutral 4-unit charge Coulomb systems must be reviewed and its implications on the computational side reassessed.

*Combining all molecular and atomic evidence collected above, there is only one conclusion possible: the existence of natural antihydrogen H̲ as well as of HH̲ in nature is established beyond doubt. Contrary to common belief and expectation, H and H̲ do not annihilate but, instead, they form a very stable bond, called the hydrogen molecule HH̲ but conventionally denoted by $H_2$ or H(↑ )H(↓).*

The price to pay for a generic solution for both H̲ and HH̲ is to admit that, historically, something in the early days of atomic, molecular quantum physics and/or theoretical chemistry has gone wrong. Bohr, Heitler and London can hardly be blamed, as the concept of antiparticles (charge-inversion) did not exist at their time. The importance of work by Hund and Lamb for chiral behavior of hydrogenic systems can hardly be overestimated. More consequences, applications and prospects are in [6,11-13,27], those relating to the ongoing H̲-experiments at CERN [5] are given elsewhere [8].

*We can safely conclude that solving some basic problems with H̲ and HH̲ in (the advent of) Physics or Einstein Year 2005 is feasible.*

**Acknowledgements.** I am in debt to a referee for constructive remarks, to Marco Tomaselli (GSI) for discussions and to Ivo Verhaeghe (U Ghent) for technical assistance.

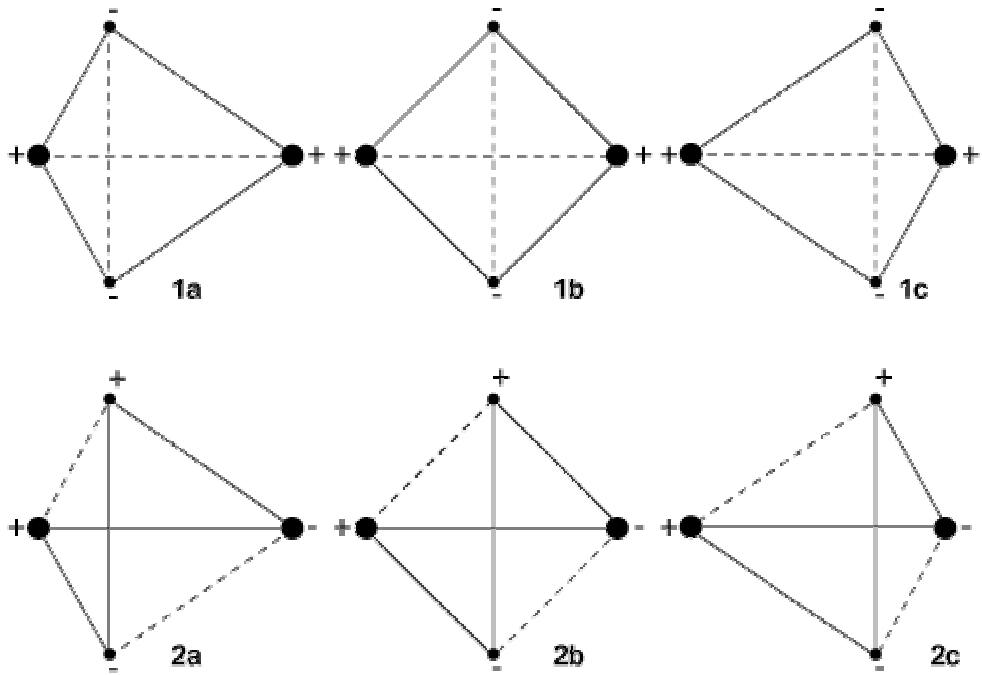

Fig. 1 The 6 Coulomb terms in (1), pseudo-ionic (a,c), covalent (b), constraint (8b) *not applicable*
Fig. 2 The 6 Coulomb terms in (3), ionic (a,c), covalent (b), constraint (8b) *applicable*
   (equatorial: baryons, axial leptons; full lines: attraction, dashed lines: repulsion)

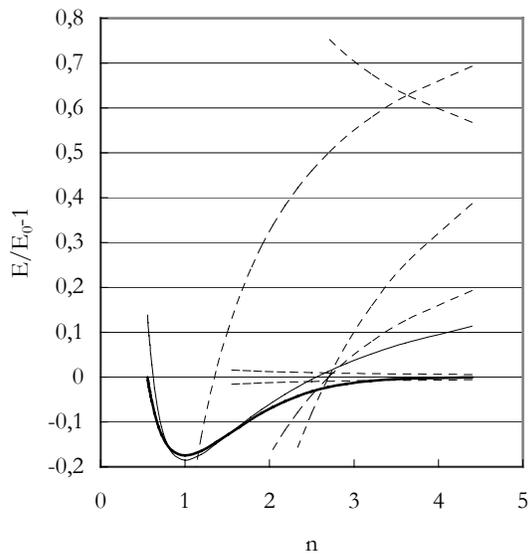

Fig. 3a Potentials (4a-d) and observed PEC for $H_2$ versus n
(from top to bottom at n=4, numbers in Table 1, dashed lines: 4(+), 1(-), 2(-), 3(-), 4(-)
full line normal 4(-), dashed lines 5(+ and -); full line bold 6)

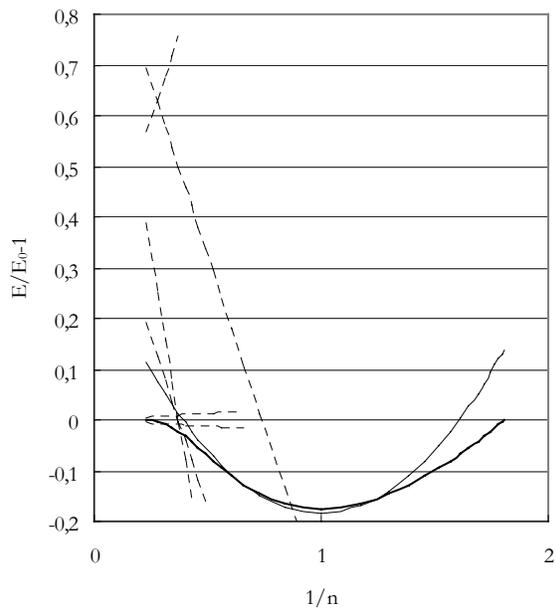

Fig. 3b Potentials (4a-d) (dashes) and observed PEC for $H_2$ (bold) versus $1/n$
(same sequence as in Fig. 3a at $1/n=0,25$)

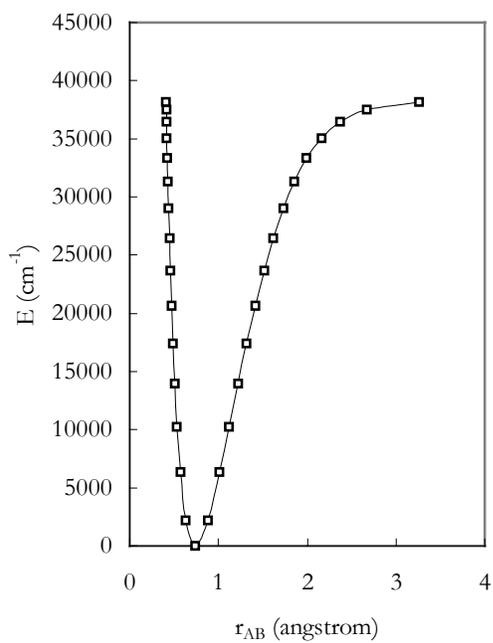

Fig. 4a Classical representation of RKR PEC for $H_2$ [19]
(□ turning points, line aid to the eye)

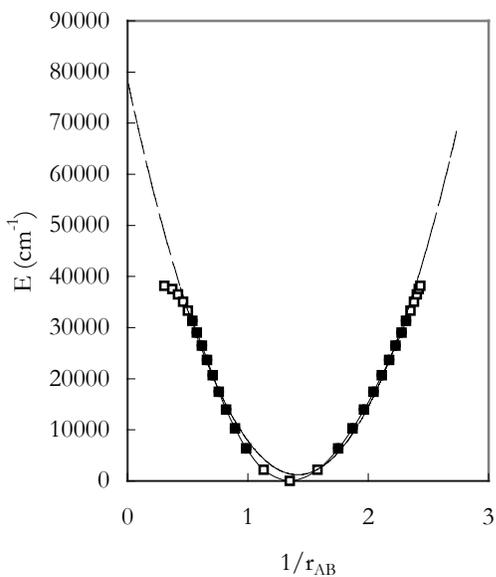

Fig. 4b Inverse presentation of RKR PEC for $H_2$
(□ all turning points as in 3b, ● turning points used for fit,
dashed line: 2$^{nd}$ order fit)

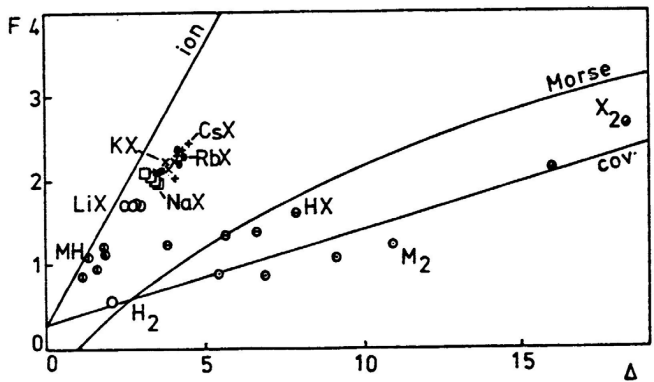

Fig. 1a. Plot of $F$ versus the Sutherland parameter $\Delta$. Solid lines represent theoretical predictions (ion, Morse) and an empirical relation (cov). The following symbols have been used: ○ $H_2$, ⊙ $M_2$, ◐ $X_2$, ⊕ HX, ⊖ MH, ○ LiX, □ NaX, × KX, ● RbX, + CsX.

Fig. 5a. Varshni function F versus covalent Sutherland parameter (11) for 39 diatomic bonds
(reproduced with permission from Verlag Zeitschrift für Naturforschung)

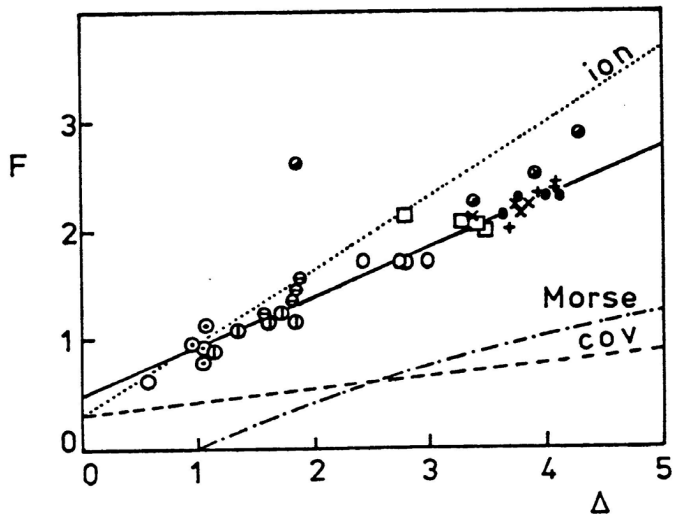

Fig. 4a. Plot of $F$ versus the "universal" Sutherland parameter $\Delta$. Same notation as in Fig. 1a.

Fig. 5b. Varshni function F versus ionic Sutherland parameter (12) for 39 diatomic bonds
(reproduced with permission from Verlag Zeitschrift für Naturforschung)

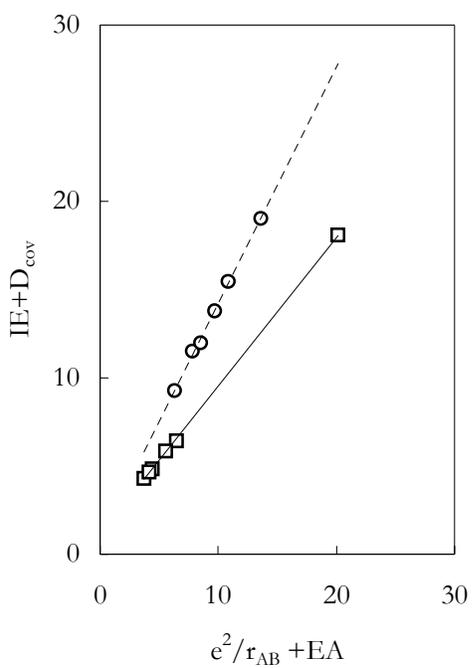

Fig. 6 Linear relation (13) between atomic and molecular constants for 12 homonuclear covalent diatomic bonds $X_2$
($\square$ H, Li, Na, K, Rb, Cs; o F, Cl, Br, I, Au, Ag)